%plain tex file of paper on spectral flow. 28 Jan 2002.

\magnification=1200
\baselineskip 14pt

\centerline{\bf SPECTRAL FLOW AND LEVEL SPACING OF EDGE STATES} 
\vskip 0.1cm
\centerline{\bf FOR QUANTUM HALL HAMILTONIANS}

\vskip 2cm

\centerline{Nicolas Macris}

\vskip 0.25cm

\centerline{Institute for Theoretical Physics}
\centerline{Ecole Polytechnique F\'ed\'erale de Lausanne}
\centerline{CH-1015 Lausanne, Switzerland}

\vskip 2cm

\centerline{\bf Abstract}

\vskip 0.25cm

\noindent We consider a non relativistic particle on the surface of a semi-infinite cylinder
of circumference $L$ submitted to a perpendicular magnetic field of strength $B$ and to the 
 potential of impurities of maximal amplitude $w$. This model is of importance 
in the context 
of the integer quantum Hall effect. In the regime of strong magnetic field or weak disorder
$B>>w$  it
is known that there are chiral edge states,
which are localised within a few magnetic lengths close to, and extended along the boundary of 
the cylinder, and whose energy levels lie in the gaps of the bulk system. 
These energy levels have
a spectral flow, uniform in $L$, as a function of a magnetic flux which 
threads the cylinder along its axis. Through a detailed study of this spectral flow we prove
that the spacing between two consecutive levels of edge states is bounded below by $2\pi\alpha L^{-1}$
with $\alpha>0$, independent of $L$, and of the configuration of impurities. This implies that the level repulsion of the chiral edge states
is much stronger than 
that of extended states in the usual Anderson model and their statistics
cannot obey one of the Gaussian ensembles. Our analysis uses the notion of relative index between two
projections and indicates that the level repulsion is connected to topological
aspects of quantum Hall systems.

\beginsection{1. INTRODUCTION AND RESULTS}

Recently there has been  mathematical progress concerning the  
spectral properties of disordered quantum Hall systems with boundaries. In the
theory of the integer quantum Hall effect one considers non-interacting electrons confined on the surface of a finite cylinder [1] or on a corbino disk [2], submitted to
a perpendicular uniform magnetic field of strength $B$ and to the
potential of impurities of maximal amplitude $w$. In a classic paper on the subject [2] Halperin argued that, at least for strong magnetic field and weak disorder
($B>>w$ in appropriate units), there  exist quantum mechanical 
states localised near and extended along the boundaries of 
the sample. These states carry a diamagnetic current 
contributing to the total Hall current. Halperin's analysis 
applies to energies that lie in the gaps separating the 
Landau bands of the bulk disordered hamiltonian, i.e the hamiltonian 
of an infinite two dimensional planar system (with no boundaries). 
Here we will call this part of the spectrum the "pure edge spectrum". 
Progress towards the characterisation of the nature of the pure edge spectrum has been made in recent works  for systems with one smooth boundary [3], [4], [5]. In the present contribution we obtain new results for such systems, which are used in separate work on more realistic geometries involving two boundaries [6].
  
We consider the Hamiltonian of a particle on a cylinder of radius ${L\over 2\pi}$ thread by a flux line with flux $\Phi$
$$
H(\Phi)={1\over 2}p_x^2+{1\over 2}(p_y-Bx+{\Phi\over L})^2 +W(x) +V(x,y)
\eqno(1.1)
$$
where $x\in {\bf R}$, $-{L\over 2}\leq y\leq {L\over 2}$, with periodic boundary conditions in the $y$ direction $\Psi(x,-{L\over 2})=\Psi(x, {L\over 2})$.
The particle is confined to the left half of the cylinder because of the external potential $W$ which models the
boundary of a "semi-infinite cylinder". We assume that it is continuous, and 
$W(x)=0$ for $x\leq 0$, $W^\prime(x)>0$ for $x\geq 0$,  
$W(x)\to +\infty$, $x\to +\infty$.  For technical reasons we  assume a growth of $W$ that is not too fast: we suppose that for $x\geq 0$, 
$u_1x^\gamma\leq W(x)\leq u_2 x^\gamma$, for some $0<u_1<u_2$ 
and $\gamma\geq 2$.
 The potential of impurities $V$ is piecewise continuous and bounded 
$|V(x,y)|\leq w$ with $0< w< {B\over 2}$.
We also suppose
that $V(x,y)=0$ for $x>0$, however our methods can be adapted to
a more general model where the impurity potential extends inside the region of the boundary.

We will also use two other Hamiltonians: the "edge hamiltonian"
$H_e(\Phi)$ obtained from (1.1) by removing $V$ and "the bulk hamiltonian" $H_b(\Phi)$ obtained from (1.1) by removing $W$.

The "semi-infinite planar" case corresponds to $L=+\infty$. In this limit the corresponding Hamiltonians become independent of $\Phi$ and we denote them $H_\infty$, $H_{e,\infty}$, $H_{b,\infty}$. It is easy to see that 
$H_{b,\infty}$ has gaps $G_n\supset ](n+{1\over 2})B+w, (n+{3\over 2})B-w[$, $n\in {\bf N}$. A basic fact
is that for weak enough disorder the "pure edge spectrum" $\sigma(H_\infty)\cap G_n$, $n\in {\bf N}$ is continuous. This result is also proven for $W$ replaced by a Dirichlet boundary condition at $x=0$ and for smooth curved open boundaries (see [3,4,5]).

When $L$ is finite $G_n$ contains only discrete isolated eigenvalues. We formulate this result and all the subsequent ones in the special case $n=0$. 

\vskip 0.5cm

\noindent{\bf Lemma 1.}
 Let $B>2w$. For any $0<\epsilon<{B\over 2}-w$ the set 
 $\sigma(H(\Phi))\cap \tilde G_0$, $\tilde G_0=]{B\over 2}+w+\epsilon,
 {3B\over 2}-w-\epsilon[$ contains only a finite number of 
 isolated eigenvalues of finite multiplicity. We label the eigenvalues of $H(0)$ in $\tilde G_0$ 
  as $E_1(0)\leq E_{2}(0)\leq...\leq E_{N}(0)$ for some finite $N$. Any 
  $E_k(0)\in \tilde G_0$ can be continued into one or several analytic
  branches $E_k(\Phi)$ for $\Phi\in 
 [0,\Phi_k]$ for some small enough $\Phi_k>0$.

\vskip 0.5cm

The discreteness of the spectrum in the specified interval is non trivial even if the circumference of the cylinder is finite because the impurity potential can extend to infinity in the direction 
$x\to -\infty$ where there is no confinement. In fact one can see that the rest of the spectrum may have  dense parts. For example if $V$ is a typical 
realisation of a random potential the Landau bands $[(n+{1\over 2})B-w,
(n+{1\over 2})B+w]$ have dense spectrum. Now let $0<\delta<{B\over 2}-w-\epsilon$ and 
$\Delta=]B-\delta, B+\delta[$. For $L$ large enough, as long as an eigenvalue
$E_k(\Phi)\in\Delta$ for some $\Phi$, then we are assured that it can be continued into an analytic branch for the whole interval $[0,2\pi]$. This comes from the fact (see inequality (3.15)) that the maximal variation of $E_k(\Phi)$
is $2\pi\sqrt{3B}L^{-1}$ so that it stays in $\tilde G_0$ and never merges in the Landau bands.

In the rest of this work we fix $\epsilon$ small and  $0<\delta<{B\over 2}-w-\epsilon$, and look only at eigenvalues $E_k(\Phi)\in \Delta$. Note that as $\Phi$ varies from $0$ to $2\pi$
some of the branches may move in or out of $\Delta$. 
A reformulation of the analysis in [3,4,5] shows that there exists a spectral flow which 
is uniform in $L$. This is expressed by the following Lemma. 

\vskip 0.5cm

\noindent {\bf Lemma 2.}
 Let $B>2w$. There exists $\delta$, $ w_0$ small enough, $ L_0$ large enough such that for $w< w_0$, $L> L_0$ all eigenvalues $E_k(\Phi)\in\Delta$ satisfy
$$
L{d\over d\Phi}E_k(\Phi)\geq \alpha
\eqno(1.2)
$$
where $\alpha$ is strictly positive independent of $L$ and $k$, and 
depends only on $W$, $B$, $w$ and $\delta$.

\vskip 0.5cm

The existence of a spectral flow is equivalent to the presence of a chiral diamagnetic
current. Indeed by the Feynman-Hellman theorem
$$
{d\over d\Phi}E_k(\Phi)=j_k(\Phi)
\eqno(1.3)
$$
where 
$$
j_k(\Phi)={1\over L}<\Psi_k(\Phi)|(p_y-Bx+{\Phi\over L})\Psi_k(\Phi)>
\eqno(1.4)
$$
is the diamagnetic current (or edge current) associated to the eigenstate
$|\Psi_k(\Phi)>$ corresponding to the level $E_k(\Phi)$.

The hamiltonians $H(\Phi)$ and $H(\Phi+2\pi)$ are unitarily equivalent, the unitary
operator being multiplication by $\exp(2\pi i{y\over L})$. Thus for each $E_k(\Phi)$ which does not merge in the Landau bands there must exist some $k^\prime$ such that $E_k(2\pi)=E_{k^\prime}(0)$. From
Lemma 2 it is clear that $k^\prime>k$, but this does not characterise completely the spectral flow.
Our main new result states that $k^\prime=k+1$ and  characterises  the level spacing for the pure edge spectrum.

\vskip 0.5cm

\noindent {\bf Theorem 1.}
Let $B>2w$. There exist $\delta$,  $w_0$ small enough, $L_0$ large
enough such that for $w< w_0$, $L> L_0$, the branches $E_k(\Phi)$ belonging to $\Delta$ for all $\Phi\in [0,2\pi]$ satisfy
$$
E_k(2\pi)=E_{k+1}(0)
\eqno(1.5)
$$
Moreover  the level spacing in $\Delta$
satisfies
$$
{2\pi\alpha\over L}\leq |E_{k+1}(0)-E_k(0)|\leq {2\pi\sqrt{3B}\over L}
\eqno(1.6)
$$

\vskip 0.5cm

For the constant $\alpha$ in Lemma 2 and 
theorem 1 we can take the right hand side of (2.29). The important point is that in the lower bound of (1.6) $\alpha$ does not depend on the detailled configuration of the impurity
potential but only on its maximal amplitude. So for a random potential the 
level spacing is random but our lower bound is non random.

\vskip 0.5cm

For the usual Anderson model it is proven that the level spacing of localised
states satisfies Poisson statistics [7], [8] and it is numerically established that 
 extended states have a level repulsion satisfying the Wigner surmise [9]. 
Here we have a different situation: the states are extended, chiral 
and have a much stronger level repulsion which makes the level spacing very rigid. 
Let $\rho(E)$ denote the average density of edge states.
We expect from (1.6) 
that, in the limit $L\to\infty$, the rescaled level spacing $s=L\rho(E_k)|E_{k+1}-E_k|$
 has a histogram $p(s)$ which is a certain broadening of 
$\delta(s-1)$ with a finite support of $O({w^2\over B^2})$. The level
statistics cannot follow the Gaussian ensembles and it would be worthwhile to investigate this question numericaly for 
an analogous model on a lattice. It is apparent from the proof of theorem 1 that the rigidity of the edge spectrum
is related to the topological invariants of the quantum Hall effect. Also if
the spectral flow would satisfy $E_k(2\pi)=E_{k+n}(0)$ with $n\geq 2$, it
would not be forbidden to have $n$ consecutive levels arbitrarily close.

We wish to point out that all these features can be checked immediately for a simple toy Hamiltonian. Consider a one dimensional chiral
particle on a circle of circumference $L$ thread by a flux $\Phi$
$$
h(\Phi)=(-i\partial_y+{\Phi\over L}) + v(y)
\eqno(1.7)
$$
The exact spectrum is
$$
e_m(\Phi)={2\pi m\over L}+{\Phi\over L}+{1\over L}
\int_{-{L\over 2}}^{L\over 2} dy v(y)
\eqno(1.8)
$$
which satisfies (1.2), (1.5), (1.6) and has $p(s)=\delta(s-1)$. It is expected that (1.7) is a 
good approximation of (1.1) for distances to the boundary of the order 
of the magnetic length $x= O({1\over \sqrt B})$.

Finaly we recall how it follows from (1.5)  that the "edge conductance" 
of the semi-infinite system is quantized (see [1], [2], [4] for similar discussions).  
Let
$P_\Delta(\Phi)$ be the projector of $H(\Phi)$ on an energy range $\Delta$. 
The edge conductance may be defined as the total edge current per unit energy,
$$
\sigma_e=\lim_{L\to\infty}{1\over |\Delta|L}{\rm Tr}(p_y-Bx-{\Phi\over L})P_\Delta(\Phi)
\eqno(1.9)
$$
We assume that for a suitable class of potentials $V$ 
this limit exists and is independent of $\Phi$ (the flux has no effect for
the semi-infinite plane). We expect this assumption to be true 
  for typical realisations of random potentials that are ergodic with respect to the 
  translations along $y$. In this case the limit should be equal to
  ${1\over \Delta}{\rm Av}\int dx<x,0|(p_y-Bx)P_{\infty,\Delta}|x,0>$
  where ${\rm Av}$ is the average over the disorder and $P_{\infty, \Delta}$
the projector of $H_\infty$ onto $\Delta$. The limit of the later
  quantity when $\Delta\to \mu$ has been shown to be an integer if 
  $\mu$ is a point in the gap $G_0$, by 
  non-commutative geometry techniques applied to the lattice case [10].
 In the present situation it is easy to see that for $\Delta$ in the first gap of the bulk Hamiltonian
$H_{b,\infty}$
$$
\eqalign{   
{1\over |\Delta|L} & ||(p_y-Bx-{\Phi\over L})P_\Delta(\Phi)||_1 
\leq
{1\over |\Delta|L}||(p_y-Bx-{\Phi\over L})P_\Delta(\Phi)||.||P_\Delta(\Phi)||_1 
\cr &
\leq {\sqrt 2\over |\Delta|L}{\rm sup}_{||\psi||=1}
(<\psi|P_\Delta(\Phi)(H(\Phi)-V)P_\Delta(\Phi)|\psi>)^{1/2} {\rm Tr} P_\Delta(\Phi) 
\cr &
\leq  {\sqrt{3B}\over |\Delta|L}{\rm Tr} P_\Delta(\Phi)=O(1)
\cr}
\eqno(1.10)
$$
 Here
$||.||_1$ and $||.||$ are the trace and operator norms respectively and we used$||AB||_1\leq ||A||.||B||_1$ for $A$ bounded and $B$ trace class. In the last equality we used
that there are $O(L)$ states in $\Delta$ because of (1.6) so that the final bound is uniform 
with respect to $L$.
Since we have assumed that $\sigma_e$ is independent of $\Phi$, by averaging over $\Phi$
we get
$$
\eqalign{
\sigma_e & = \lim_{L\to\infty}{1\over|\Delta|}\int_0^{2\pi}{d\Phi\over 2\pi}
\sum_{E_k(\Phi)\in\Delta}{dE_k(\Phi)\over d\Phi} =
\lim_{L\to\infty}
{1\over |\Delta|}\sum_{k_{min}}^{k_{max}} \int_0^{2\pi} {d\Phi\over 2\pi}{dE_k(\Phi)\over d\Phi}
\cr &
=\lim_{L\to\infty}
{1\over 2\pi |\Delta|}\sum_{k_{min}}^{k_{max}} (E_{k+1}(0)-E_k(0))  
=\lim_{L\to\infty}{1\over 2\pi|\Delta|}(E_{k_{max}}-E_{k_{min}}) ={1\over 2\pi}
\cr}
\eqno(1.11)
$$
For the first equality we use (1.3), (1.4) and dominated convergence.
To obtain the second equality we consider separately the contributions of the eigenvalues with $k_{min}\leq k\leq k_{max}$ such that $E_k(\Phi)\in \Delta$
for all $\Phi\in [0, 2\pi]$, and of a finite number of eigenvalues with $k<k_{min}$ (resp. $k>k_{max}$) which enter (resp. leave) $\Delta$ as $\Phi$ varies from $0$ to $2\pi$. From (1.6) and (3.15) this later contribution is $O(L^{-1})$.
Finaly (1.5) is used in the third equality. Here
the units are such that $e=\hbar=1$ so ${1\over 2\pi}={e^2\over h}$.

Section 2 contains the proofs of Lemmas 1 and 2 and a third Lemma that 
is needed for the proof of theorem 1 in section 3. 
The appendices A and B contain technical estimates.
 
\vskip 0.5cm 

\beginsection{2. DISCRETENESS OF EDGE SPECTRUM AND SPECTRAL FLOW} 

\vskip 0.5cm

\noindent {\bf Proof of Lemma 1}
 
Let $D>0$ to be chosen later (large) and $V_D(x,y)=V(x,y)$
 for $x\leq -D$, $V_D(x,y)=0$ for $x> -D$. Then $V(x,y)-V_D(x,y)$ has compact support and a standard argument using the resolvent identity implies that the essential spectra of 
$$
H_D(\Phi)=H_e(\Phi)+V_D(x,y)
\eqno(2.1)
$$
and 
$$
H(\Phi)=H_D(\Phi)+V(x,y)-V_D(x,y)
\eqno(2.2)
$$
coincide [11]. Therefore if we show that $\sigma(H_D(\Phi))\cap \tilde G_0$ contains 
only isolated eigenvalues of finite multiplicity, the same is true for $H(\Phi)$. This 
will be achieved below using a decoupling scheme [12], [13] which proves 
that $\sigma(H_D(\Phi))\cap\tilde G_0$ is a small perturbation of $\sigma(H_e(\Phi))\cap\tilde G_0$.
The set $\sigma(H_e(\Phi))$ consists of non degenerate energy levels 
$\epsilon_n({2\pi m\over L}+{\Phi\over L})$, $n\in {\bf N}$ the Landau 
index and $m\in {\bf Z}$, where $\epsilon_n(k)$, $k\in {\bf R}$ the wavenumber 
conjugate to $y$, are the spectral branches of $H_{e,\infty}$. These spectral 
branches are monotone increasing entire functions of $k$ with $\epsilon_n(k)\to +\infty$ 
for $k\to+\infty$ and $\epsilon_n(k)\to (n+{1\over 2})B$ for $k\to -\infty$
(see for example [3]). 

In order to set up the decoupling scheme we introduce the characteristic functions $\chi_e(x)$ of $-{D\over 2}\leq x<+\infty$ and $\chi_b(x)$ of 
$-\infty\leq x<-{D\over 2}$. Note that $\chi_e(x)+\chi_b(x)=1$ for all $x$. We also need the monotone and twice differentiable functions 
$J_e(x)$, $J_b(x)$ such that 
$J_e(x)=0$ for $-\infty<x<-{3D\over 4}-1$ and
$J_e(x)=1$ for $-{3D\over 4}+1<x<\infty$; 
$J_b(x)=1$ for $-\infty<x<-{D\over 4}-1$, $J_b(x)=0$ for $-{D\over 4}+1<x<\infty$.

We introduce the Green functions $G_\alpha(z)=(H_\alpha(\Phi)-z)^{-1}$ for $\alpha=e, b, D$ and $z\in {\bf C}$ in the resolvent set of the corresponding hamiltonian. Since
$$
H_D(\Phi)J_\alpha=H_\alpha(\Phi)J_\alpha\qquad {\rm for}\qquad \alpha=e,b
\eqno(2.4)
$$
following [13] we have
$$
\eqalign{
(H_D(\Phi)-z)&(J_eG_e(z)\chi_e+J_bG_b(z)\chi_b)
\cr &=(H_e(\Phi)-z)J_eG_e(z)\chi_e+(H_b(\Phi)-z)J_bG_b(z)\chi_b
\cr &
= J_e\chi_e+J_b\chi_b+{1\over 2}[p_x^2, J_e]G_e(z)\chi_e
+{1\over 2}[p_x^2, J_b]G_b(z)\chi_b
\cr &
=1+K_e(z)+K_b(z)
\cr}
\eqno(2.5)
$$
 where $K_\alpha(z)={1\over 2}[p_x^2, J_\alpha]G_\alpha(z)\chi_\alpha$, $\alpha=e, b$. Thus 
$$
(H_D(\Phi)-z)^{-1}=(J_eG_e(z)\chi_e+J_bG_b(z)\chi_b)(1+K_e(z)+K_b(z))^{-1}
\eqno(2.6)
$$
In Appendix A we prove the following estimates for the operator norms of 
$K_e(z)$ and $K_b(z)$ for 
${B\over 2}+w<{\rm Re} z<{3B\over 2}-w$ (in what follows $c$ is a generic positive numerical constant)
$$
||K_e(z)||\leq{c B^{3\over 2}L\over \delta_e(z)}e^{-cBD^2}
\eqno(2.7)
$$
$$
||K_b(z)||\leq{cB^{3\over 2}L \over \delta_0(z)-cw}e^{-c \sqrt B D}
\eqno(2.8)
$$
where 
$\delta_e(z)={\rm dist}(z, \sigma(H_e(\Phi))$ and where
$\delta_0(z)=\min(|z-{B\over 2}|, |z-{3B\over 2}|)$. We have to take $w$
small enough so that the denominator in (2.8) stays positive. Later on we choose $z$ appropriately and $D$ large enough so that both terms become smaller than ${1\over 2}$. Thus
$$
(H_D(\Phi)-z)^{-1}=J_eG_e(z)\chi_e+J_bG_b(z)\chi_b +R(z)
\eqno(2.9)
$$
where 
$$
||R(z)||\leq \biggl(||G_e(z)||+||G_b(z)||\biggr)\biggl[(1-||K_e(z)||-||K_b(z)||)^{-1}-1\biggr]
\eqno(2.10)
$$
Let $m\in {\bf Z}$ be such that $\epsilon_0({2\pi m\over L}+{\Phi\over L})$
is an eigenvalue belonging to $\sigma(H_e(\Phi))\cap \tilde G_0$. We can choose 
$\rho>0$ small enough independent of $m$ and $L$ such that the circle $C_m$ with center $\epsilon_0({2\pi m\over L}+{\Phi\over L})$ and radius ${\rho\over L}$
encloses only one such eigenvalue. By choosing $z$ in a sufficiently thin annulus around $C_m$ and $D$ large enough, (2.7) and (2.8) can be made smaller than
${cB^{3\over 2}L^2\over \rho}e^{-c\sqrt B D}<{1\over 2}$. At the same time from (2.10) we have
$$
||R(z)||\leq {cB^{3\over 2}L^3\over \rho^2}e^{-c\sqrt B D}
\eqno(2.11)
$$
so that from (2.9) $(H_D(\Phi)-z)^{-1}$ is well defined for $z$ in a thin annulus surrounding $C_m$. Therefore we can compute the spectral projection $P_D(m, \Phi)$ of $H_D(\Phi)$ for the interval 
$I_m=]\epsilon_0({2\pi m\over L}+{\Phi\over L})-{\rho\over L}, 
\epsilon_0({2\pi m\over L}+{\Phi\over L})+{\rho\over L}[$ by Cauchy's formula.
Let $P_e(m, \Phi)$ be the projector of $H_e(\Phi)$ corresponding to the level $\epsilon_0({2\pi m\over L}+{\Phi\over L})$. Thanks to (2.9), (2.11) we obtain for $D$ large enough
$$
||P_D(m, \Phi)-P_e(m, \Phi)||\leq {cB^{3\over 2}L^2\over \rho}e^{-c\sqrt B D}<1
\eqno(2.12)
$$
This estimate implies that $\sigma(H_D(\Phi))\cap I_m$ contains only one eigenvalue of multiplicity equal to one. Note that this conclusion holds for all 
$I_m\subset \tilde G_0$. Finally since $H_e(\Phi)$ and $H_b(\Phi)$
have no spectrum in $(\cup_m I_m)^{c}\cap \tilde G_0$ we deduce from (2.7), (2.8), (2.9)
that $H_D(\Phi, L)$ has no spectrum in that same set. Therefore 
$\sigma(H_D(\Phi, L))\cap \tilde G_0$ consists of isolated eigenvalues of multiplicity one.

It remains to show that an eigenvalue 
$E_k(0)\in \tilde G_0$  can be continued into one or several analytic branches $E_k(\Phi)$ for $\Phi$ small enough.
In the present case it is sufficient to show [11] that 
$(py-Bx)$ is relatively bounded with respect to
$H(0)$. For any $\psi$ in the domain of 
$H(0)$ and any complex number $z$ with ${\rm Im}z\neq 0$ we have 
$$
\eqalign{{1\over 2}
||&(p_y-Bx)\psi||^2  \leq <\psi|(H(0)-V)\psi>
\cr &
=<\psi|(H(0)-z)^{-1}(H(0)-z)|(H(0)-\bar z+z)\psi>  -<\psi|V\psi>
\cr &
\leq  ||(H(0)-z)^{-1}||.||H(0)\psi||^2 +|z|.||\psi||^2+|z|^2||(H(0)-z)^{-1}||.||\psi||^2
+w||\psi||^2
\cr &
\leq 
{1\over |{\rm Im} z|} ||H(0)\psi||^2+(|z|+{|z|^2\over |{\rm Im} z|}+w)||\psi||^2
\cr}
\eqno(2.13)
$$
This concludes the proof of the Lemma.

\vskip 0.5cm

\noindent{\bf Remark:} In (2.13) we can take $|{\rm Im} z|$ as large as we wish so the size of the interval of analyticity is not limited by the relative bound but rather by the fact that the branch $E_k(\Phi)$ may merge in the Landau bands
(outside of $G_0$) where it may not be isolated anymore. Inequality (3.15) shows that for $L$ large enough the maximal variation of $E_k(\Phi)$ is $2\pi\sqrt{3B}L^{-1}$, so that if $E_k(\Phi)$ is contained in $\Delta$ for some $\Phi$ then it is contained in $\tilde G_0$ and it is analytic
for all $\Phi\in [0,2\pi]$.

\vskip 0.5cm

Before presenting the formal proof of Lemma 2 we would like to point out that in fact (1.2) is closely related to the ideas in [3] and [4]. Using the unitary translation operator $x\to x+{\Phi\over BL}$ and the Feynman-Hellman theorem it is easy to see that
$$
L{d\over d\Phi}E_k(\Phi)=<\Psi_k(\Phi)|(W^\prime+\partial_x V)\Psi_k(\Phi)>
$$
where $|\Psi_k(\Phi)>$ is the eigenstate with eigenvalue $E_k(\Phi)$. Using 
the methods of [3] or [4] one may show that for $E_k(\Phi)\in \Delta$,
$|\Psi_k(\Phi)>$ is mainly concentrated near the region where $W^\prime(x)$ is large so 
that (1.2) holds provided both $V$, $\partial_x V$ are small enough. Here 
we follow a different method which is closer to the original argument of Halperin [2] in that 
it uses directly the relation (1.4) instead of (2.13). Only the smallness of $V$ is required.

\vskip 0.5cm

\noindent{\bf Proof of Lemma 2}

The eigenstates $|u_{nm}(\Phi)>$ of $H_e(\Phi)$ with eigenvalues 
$\epsilon_n({2\pi m\over L}+{\Phi\over L})$ are of the form
$$
<xy|u_{nm}(\Phi)> = e^{i{2\pi m\over L}y} h_{nm}(x)
\eqno(2.14)
$$
so that $<u_{nm}(\Phi)|(p_y-Bx-{\Phi\over L})u_{n^\prime m^\prime}(\Phi)>=0$
for $m\neq m^\prime$ and all $n$, $n^\prime$. Therefore writing
$$
|\Psi_k(\Phi)>=|\Psi_k^0(\Phi)>+|\Psi_k^1(\Phi)>
\eqno(2.15)
$$
where 
$$
|\Psi_k^0(\Phi)>=\sum_{m=-\infty}^{+\infty} c_{k}^{0m}|u_{0m}(\Phi)>
\eqno(2.16)
$$
$$
|\Psi_k^1(\Phi)>=\sum_{n\geq 1}\sum_{m=-\infty}^{+\infty} c_{k}^{nm}|
u_{nm}(\Phi)>
\eqno(2.17)
$$
we obtain from (1.3), (1.4)
$$
\eqalign{
L{d\over d\Phi}E_k(\Phi)=&
\sum_{m=-\infty}^{+\infty} |c_{k}^{0m}|^2<u_{0m}(\Phi)|(p_y-Bx-{\Phi\over L})u_{0m}(\Phi)>
\cr &
+2{\rm Re}<\Psi_k^0(\Phi)|(p_y-Bx-{\Phi\over L})\Psi_k^1(\Phi)>
\cr &
+<\Psi_k^1(\Phi)|(p_y-Bx-{\Phi\over L})\Psi_k^1(\Phi)>
\cr}\eqno(2.18)
$$
First we show that the last two terms on the right hand side of (2.18) are bounded by the norm $\sqrt{3B}||\Psi_k^1(\Phi)||$. The Schwartz inequality implies
$$
\eqalign{
|<\Psi_k^0(\Phi)| & (p_y-Bx-{\Phi\over L})\Psi_k^1(\Phi)>| 
\leq ||\Psi_k^1(\Phi)||.||(p_y-Bx+{\Phi\over L})\Psi_k^0(\Phi)||
\cr &
\leq\sqrt{2}||\Psi_k^1(\Phi)||
(<\Psi_k^0(\Phi)|H_e(\Phi)\Psi_k^0(\Phi)>)^{1/2}
\cr &
\leq
\sqrt{2}||\Psi_k^1(\Phi)||
\biggl(<\Psi_k^0(\Phi)|H_e(\Phi)\Psi_k^0(\Phi)>
+ \cr & <\Psi_k^1(\Phi)|H_e(\Phi)\Psi_k^1(\Phi)>\biggr)^{1/2}
\cr &
=
\sqrt{2}||\Psi_k^1(\Phi)||
(<\Psi_k(\Phi)|H_e(\Phi)\Psi_k(\Phi)>)^{1/2}
\cr &
\leq
\sqrt{2}||\Psi_k^1(\Phi)||\bigl(E_k(\Phi)+w\bigr)^{1/2}
\leq \sqrt{3B}||\Psi_k^1(\Phi)||
\cr}
\eqno(2.19)
$$
For the third matrix element on the right hand side of (2.18) the same method leads to an identical estimate.
From the Feynman-Hellman formula we have
$$
\eqalign{
<u_{0m}(\Phi)|(p_y-Bx-{\Phi\over L})u_{0m}(\Phi)> &
=L{d\over d\Phi}\epsilon_0({2\pi m\over L}+{\Phi\over L})
\cr &
=
\epsilon_0^{\prime}({2\pi m\over L}+{\Phi\over L})
\cr}
\eqno(2.20)
$$
where $\epsilon_0^\prime(k)$ is the derivative of the lowest monotone increasing spectral branch corresponding to the hamiltonian $H_{e,\infty}$. From (2.18), (2.19), (2.20)
$$
\eqalign{
L{d\over d\Phi}E_k(\Phi) & \geq
\sum_{m=-\infty}^{+\infty}|c_k^{0m}|^2\epsilon_0^{\prime}({2\pi m\over L}+{\Phi\over L})-2\sqrt{3B}||\Psi_k^1(\Phi)||
\cr &
\geq
v_F(M)
\sum_{|m-M|\leq \bar m}|c_k^{0m}|^2
-2
\sqrt{3B}||\Psi_k^1(\Phi)||
\cr}
\eqno(2.21)
$$
with the Fermi velocity 
$$
v_F(M)={\rm min}_{|m-M|\leq \bar m}\epsilon_0^\prime({2\pi m\over L}+{\Phi\over L})
\eqno(2.22)
$$
The integers $M$ and $\bar m$ 
 will be choosen conveniently below. Writting the Schr\"odinger equation in the form,
$$
\sum_{n=0}^{\infty}\sum_{m=-\infty}^{+\infty}
c_k^{nm}\biggl(\epsilon_n({2\pi m\over L}+{\Phi\over L})-E_k(\Phi)\biggr)|u_{nm}(\Phi)>=V(x,y)|\Psi_k(\Phi)>
\eqno(2.23)
$$
and taking the norm on both sides
$$
\sum_{n=0}^{\infty}\sum_{m=-\infty}^{+\infty}
|c_k^{nm}|^2\biggl(\epsilon_n({2\pi m\over L}+{\Phi\over L})-E_k(\Phi)
\biggr)^2\leq
w^2
\eqno(2.24)
$$
Dropping the term $n=0$,  using 
$\bigl(\epsilon_n({2\pi m\over L}+{\Phi\over L})-E_k(\Phi)
\bigr)^2\geq ({B\over 2}-\delta)^2$ for $n\geq 1$ and $E_k(\Phi)\in \Delta$ we get
$$
\sum_{n\geq 1}\sum_{m=-\infty}^{\infty}|c_k^{nm}|^2=
||\Psi_k^1(\Phi)||^2\leq {w^2\over ({B\over 2}-\delta)^2}
\eqno(2.25)
$$
From (2.24) one can also derive a lower bound for $\sum_{|m-M|<\bar m}|c_k^{0m}|^2$. Indeed retaining only the term $n=0$ and using the monotonicity of 
$\epsilon_0({2\pi m\over L}+{\Phi\over L})$ we have
$$
A(M, \bar m)^2
\sum_{|m-M|>\bar m}|c_k^{0m}|^2\leq w^2
\eqno(2.26)
$$
where $A(M,\bar m)$ is the smallest of the two numbers 
$|\epsilon_0({2\pi\over L}(M\pm\bar m)+{\Phi\over L})|-E_k(\Phi)|$.
Now we choose any $M$ such that $\epsilon_0({2\pi M\over L}+{\Phi\over L})\in\Delta$ and since $E_k(\Phi)\in \Delta$ we can take $\bar m$ such that $A(M,\bar m)\geq {B\over 2}-2\delta$. Thus
$$
\sum_{|m-M|>\bar m}|c_k^{0m}|^2\leq {w^2\over ({B\over 2}-2\delta)^2}
\eqno(2.27)
$$
Finaly the normalisation condition for $|\Psi_k(\Phi)>$ combined with (2.25)
and (2.27) imply
$$
\sum_{|m-M|\leq\bar m}|c_k^{0m}|^2\geq 1-{2w^2\over ({B\over 2}-2\delta)^2}
\eqno(2.28)
$$
From (2.21), (2.25) and (2.28) we have
$$
L{d\over d\Phi}E_k(\Phi)
\geq v_F(M)\biggl[1-
2(1+{\sqrt{3B}\over v_F(M)}){w^2\over ({B\over 2}-2\delta)^2}\biggr]
\eqno(2.29)
$$
Clearly $v_F(M)$ is a strictly positive number which does not depend on 
$V$ but only on $W$ and $B$. Therefore (2.29) implies 
the result of the Lemma for $w$ and $\delta$ small enough.

\vskip 0.5cm

It will become clear in the next section that the proof of Theorem 1 requires
 the absence of crossings for the branches $E_k(\Phi)$ in $\Delta$.
 Since we do not know a priori if this is true for $H(\Phi)$,
 an intermediate step is to construct a suitable perturbation of $H(\Phi)$
 for which the non-crossing property is satisfied. 
The perturbation that is added here has the effect to lift the degeneracy at each crossing in $\Delta$ in a way that (1.2) still holds for the perturbed branches. 
This is the content of the next Lemma.

\vskip 0.5cm

\noindent{\bf Lemma 3.} Fix $B$, $w$, $\delta$ and $L$ as in Lemma 2.
Assume that $V(x,y)$ is such that the eigenvalues
$E_l(0)$ are not degenerate. 
One can construct a finite rank perturbation 
$R(\Phi)$ with $||R(\Phi)||\leq L^{-10}$ such that the spectrum of
$\tilde H(\Phi)
=H(\Phi)+R(\Phi)$ in $\Delta$
consists of non degenerate eigenvalues forming infinitely differentiable
spectral branches which do not 
cross and are labeled as $\tilde E_l(\Phi)$ with
 $\tilde E_l(0)= E_l(0)$. Moreover the new branches satisfy
$$
L{d\over d\Phi}\tilde E_l(\Phi)\geq \tilde\alpha
\eqno(2.30)
$$
where $\tilde \alpha$ is strictly positive and independent of $L$.

\vskip 0.5cm

\noindent{\bf Proof of Lemma 3.}

Let $P_\Delta(\Phi)$ be the eigenprojector of $H(\Phi)$ onto 
$\Delta$. Then we have
$$
P_\Delta(\Phi) H(\Phi)P_\Delta(\Phi)=\sum_{E_l(\Phi)\in \Delta} E_l(\Phi)
|\Psi_l(\Phi)><\Psi_l(\Phi)|
\eqno(2.31)
$$
Since the branches $E_l(\Phi)$ are analytic and the eigenvalues are not
degenerate for $\Phi=0$ the possible crossings are necessarily isolated. Indeed
if two branches would coincide on a set with accumulation points they would concide over the whole interval $[0,2\pi]$ and therefore violate the non degeneracy assumption at $\Phi=0$. Therefore we can assume without loss of generality that there is at most a finite number of crossings in $\Delta$.
Let us construct the perturbation $R(\Phi)$. First consider the set ${\cal C}$ of pairs of branches
 which cross in $\Delta$ (note that $n$ branches may cross at the same point
 and contribute as ${n(n-1)\over 2}$ pairs).
 Pick one pair of branches in ${\cal C}$ say $(ij)$ and assume
 $E_i(0)<E_j(0)$. Suppose they cross
at points $\Phi_{ij}^\mu$ where the label $\mu$ takes into account the fact that the branches $i$ and $j$ may cross more than once, i.e
$$
E_i(\Phi_{ij}^\mu)=E_j(\Phi_{ij}^\mu)
\eqno(2.32)
$$
Let $\lambda_{ij}^\mu(\Phi)$ be infinitely differentiable test functions centered
at $\Phi_{ij}^\mu$, with a compact support of width $\beta_1$ and $\max_{0\leq \Phi\leq 2\pi} |\lambda_{ij}^\mu(\Phi)|\leq \lambda_1$. The real numbers $\delta_1$ and 
$\lambda_1$ will be adjusted in a suitable way below. Add to the Hamiltonian $H(\Phi)$ the perturbation
$$
R_1(\Phi)=\sum_{\mu} \lambda_{ij}^\mu(\Phi)
\bigl(|\Psi_i(\Phi)><\Psi_j(\Phi)|+
|\Psi_j(\Phi)><\Psi_i(\Phi)|\bigr)
\eqno(2.33)
$$
We take $\beta_1$ small enough
so that the supports of the test functions do not contain $\Phi=0$ and do not overlap.
In order to diagonalise the new hamiltonian it is sufficient to work in the two dimensional subspace of the branches $i$ and $j$. The spectral branches of the new Hamiltonian do not change for $k\neq i,j$, whereas for $k=i,j$ they 
become 
$$
E_i^{1}(\Phi)={1\over 2}\biggl(E_i(\Phi)+E_j(\Phi)-\sqrt{
(E_i(\Phi)-E_j(\Phi))^2+\lambda_{ij}^\mu(\Phi)^2}\biggr)
\eqno(2.34)
$$
and 
$$
E_j^{1}(\Phi)={1\over 2}\biggl(E_i(\Phi)+E_j(\Phi)+\sqrt{
(E_i(\Phi)-E_j(\Phi))^2+\lambda_{ij}^\mu(\Phi)^2}\biggr)
\eqno(2.35)
$$
Since the difference 
$$
E_j^{1}(\Phi)-E_i^{1}(\Phi)=\sqrt{
(E_i(\Phi)-E_j(\Phi))^2+\lambda_{ij}^\mu(\Phi)^2}
\eqno(2.36)
$$
is always strictly positive the new pair $(ij)$ is non degenerate 
for all values of $\Phi$.
Moreover by choosing $\lambda_1$ small enough we can make sure that we do not
introduce more crossings. Therefore the perturbed hamiltonian
$$
H_1(\Phi)=H(\Phi)+ R_1(\Phi)
\eqno(2.37)
$$
has a new set ${\cal C}_1$ of pairs of branches which cross, with one element
less than ${\cal C}$. One can construct in the same way a 
perturbation $R_2(\Phi)$ of (2.37) (with $\delta_2$, $\lambda_2$ small enough) so that  the new
Hamiltonian $H_2(\Phi)=H_1(\Phi)+R_2(\Phi)$ has two less  pairs of branches which cross than $H(\Phi)$.
Since there is at most a finite number of such pairs by iterating this construction
we end up with the Hamiltonian 
$$
\tilde H(\Phi)=H(\Phi)+\sum_p R_p(\Phi)=H(\Phi)+R(\Phi)
\eqno(2.38)
$$ 
of the Lemma, where the sum over $p$ contains a finite number of terms. Note that 
$\tilde H(0)=H(0)$ so that the labelling of the Lemma holds.
The norm of the total perturbation is
$$
||R(\Phi)||\leq \sum_{p}||R_p(\Phi)||\leq 
\sum_{p}\lambda_p
\eqno(2.39)
$$
The condition $||R(\Phi)||\leq L^{-10}$ can always be achieved by 
choosing at each step 
$$
\lambda_p\leq {\beta_p\over  L^{10+p}}
\eqno(2.40)
$$ 
and 
$\beta_p\leq {1\over 10}$.

It remains to check that (2.30) holds. From the formulas (2.34), (2.35)
 and Lemma 2, it is easy to check that at the first step of the construction
the new branches have new derivatives satisfying
$$
{d\over d\Phi}E_{i,j}^{1}(\Phi)\geq
{\rm min}\bigl({d\over d\Phi}E_i(\Phi),{d\over d\Phi}E_j(\Phi)\bigr) 
- {1\over 2}|{d\over d\Phi}\lambda_{ij}^{\mu}(\Phi)|
\eqno(2.41)
$$
for all $\Phi$. At each step of the construction it is possible to choose test functions such that
$$
{\rm max}_{0\leq \Phi\leq 2\pi}|{d\over d\Phi}\lambda_{ij}^\mu(\Phi)|\leq {2
\over  L^{10+p}}
\eqno(2.42)
$$
in a way consistent with (2.40). So at the first step ($p=1$) 
$$
{d\over d\Phi}E_{i,j}^{1}(\Phi)\geq
{\alpha\over L}- {1\over L^{11}}
\eqno(2.43)
$$
Of course (2.43) is also valid for the spectral branches of $H_1(\Phi)$ that correspond to $k\in {\cal N}$. Therefore it is valid for all eigenvalues 
of $H_1(\Phi)$.
By iterating the construction we see that any branch of (2.38) satisfies
$$
{d\over d\Phi}\tilde E_{l}(\Phi)\geq
{\alpha\over L}- \sum_p{1\over L^{10+p}}
\eqno(2.44)
$$
which implies (2.30).

\vskip 0.5cm

\beginsection{3. RELATIVE INDEX AND LEVEL SPACING}

The main goal of this section is to prove Theorem 1. 
Let us first outline  the strategy of the proof. Without loss of 
generality we can suppose that $V$
is such that $E_k(0)$ are non degenerate. Indeed if this is not 
the case one may find a sufficiently small perturbation $u(x,y)$,
$||u||_\infty<L^{-10}$ such that this hypothesis is satisfied for $V+u$. If (1.5), (1.6) hold for 
$V+u$ then they hold for $V$ because the perturbation of the discrete levels separated by 
$O(L)$ is at most
$O(L^{-10})$. 
From Lemma 2 we know that for $E_k(\Phi)\in \Delta$ there is a non 
trivial spectral flow: the branches are monotone increasing and 
since $H(0)$ and $H(2\pi)$ are unitarily equivalent we must have 
$E_k(2\pi)=E_{k^\prime}(0), k^\prime> k$.
We want to show that in fact $k^\prime = k+1$. Let
$E_F$ be a single "Fermi energy" lying between two consecutive 
levels of both Hamiltonians $H_D(0)$ and $\tilde H(0)$. Define the integers
$Q_F^D$ and $\tilde Q_F$ to be the number of branches of the corresponding Hamiltonians
 which cross 
$E_F$ as $\Phi$ varies from $0$ to $2\pi$. 
We will show that $Q_F^D=\tilde Q_F=1$. We know from Lemma 3 that the branches
of $\tilde H(\Phi)$ do not have crossings, and from the proof of Lemma 1 
that the same is true for the branches of $H_D(\Phi)$. This enables us to relate $\tilde Q_F$ and 
$Q_F^D$
to the notion of relative index of a pair of projections introduced by
Avron, Seiler and Simon [14]. Then by using the fact that the Fredholm index of an operator
does not change under compact perturbations we deduce that $\tilde Q_F = Q_F^D$. By explicit computation we can check that $Q_F^D=1$ and therefore $\tilde Q_F=1$ which implies 
that $\tilde E_k(2\pi) = \tilde E_{k+1}(0)$. Since the branches of 
$\tilde H(\Phi)$ are a small perturbation of those of $H(\Phi)$ we deduce
(1.5). Estimate (1.6) is then an immediate consequence.

In order to make the paper selfcontained  we give a short summary of 
the mathematical tools used below, as developed in [14]. Let $P$ and $Q$ be orthogonal 
projections on a separable Hilbert space ${\cal H}$. The pair $(P;Q)$ 
is called Fredholm if $QP$ viewed
as a map from $P{\cal H}$ to $Q{\cal H}$ is a Fredholm operator. The relative index 
${\rm Ind}(P;Q)$ of the pair is the usual 
Fredholm index of $T=QP$, that is ${\rm dimKer}(T^\dagger T)-{\rm dimKer}(TT^\dagger)$. 
One proves that $(P;Q)$ is 
a Fredholm pair if and only if $1$ and $-1$ 
 are isolated finitely degenerate eigenvalues of $P-Q$, when they belong to the spectrum. 
Moreover  
one has ${\rm Ind }(P,Q)={\rm dimKer}(P-Q-1) - {\rm dimKer}(P-Q+1)$. 
A useful formula (we use it for $m=0$) states that if $(P-Q)^{2m+1}$ is 
trace class for some integer $m$  
then $(P;Q)$ is a Fredholm pair and ${\rm Ind}(P;Q)={\rm Tr}(P-Q)^{2n+1}$, for all ${n\geq m}$. 
A central result on which we rely is that if $(P;Q)$ and $(Q;R)$ are Fredholm pairs and either
$P-Q$ or $Q-R$  is compact then $(P;R)$ is a Fredholm pair and
$$
{\rm Ind}(P;R)={\rm Ind}(P;Q) + {\rm Ind}(Q;R)
\eqno(3.1)
$$
Finaly we note that if $(P;Q)$ is Fredholm then so is $(UPU^\dagger;UQU^\dagger)$ for any 
unitary $U$ and the relative index remains invariant. Also ${\rm Ind}(P;Q)=-{\rm Ind}(Q;P)$.

\vskip 0.5cm

\noindent{\bf Relation between $\tilde Q_F$,  $Q_F^D$ and the relative index 
of a pair of projections.}

We fix $E_F\in \Delta$ between two consecutive levels of $\tilde H(0)$ and 
$\tilde H(2\pi)$ (recall that they have the same spectrum). Let 
$\tilde P_{F,0}$ (resp. $\tilde P_{F,2\pi}$) be the projectors of $\tilde H(0)$
(resp. $\tilde H(2\pi)$) onto the energy range $]-\infty, E_F]$. We also need
the projector on levels $\tilde E_k(0)$ whose spectral branch $\tilde E_k(\Phi)$ crosses $E_F$. Namely
$$
\tilde P_{F,0}^c=\sum_{\tilde E_k(0)<E_F {\rm s.t} \tilde E_k(\Phi)
{\rm crosses} E_F}
P(\tilde E_k(0))
\eqno(3.2)
$$
where $P(\tilde E_k(0))$ is the eigenprojector of $\tilde H(0)$ corresponding
to the discrete level $\tilde E_k(0)$. Since $E_F\in \Delta$ by taking $L$ large enough we are 
assured that this sum is finite and that the branches crossing
$E_F$ remain in 
$\Delta$ for all $\Phi\in [0, 2\pi]$. 

Setting $\tilde P_{F,0}^{n.c}=\tilde P_{F,0}-\tilde P_{F,0}^c$ we have

%\noindent We introduce the following eigenprojectors:

%\vskip 0.5cm 

%\noindent i) $\tilde P_F(0)$ and $P_F^D(0)$ the projectors  of $\tilde H(0)$
%and $H_D(0)$ for energies in $]-\infty, E_F]$.

%\noindent ii) $\tilde P_F(2\pi)$ and $P_F^D(2\pi)$ the projectors  of $\tilde %H(2\pi)$
%and $H_D(2\pi)$ for energies in $]-\infty, E_F]$.

%\noindent iii) $\tilde P_F^{nc}(0)$ and $P_F^{D,nc}(0)$ the projectors  of $\t%ilde H(0)$
%and $H_D(0)$ for energies in  $]-\infty, E_F]$ and such that $\tilde E_k(\Phi)%<E_F$
%and $E_k^D(\Phi)<E_F$
%for all $0\leq \Phi\leq 2\pi$. In other words only states whose energy does 
%not cross $E_F$ are retained.

%\noindent iv) $\tilde P_F^{nc}(2\pi)$ and $P_F^{D,nc}(2\pi)$ the projectors 
% of $\tilde H(2\pi)$
%and $H_D(2\pi)$ with energy in  $]-\infty, E_F]$
%and such that $\tilde E_k(\Phi)<E_F$ and $E_k^D(\Phi)<E_F$ for all $0\leq \Phi% <2\pi$.
%Note that since the branches $E_k(\Phi)$ are monotone increasing 
%$\tilde P_F^{nc}(2\pi)=\tilde P_F(2\pi)$ and $P_F^{D, nc}(2\pi)=P_F^D(2\pi)$.

%\noindent v) $\tilde P_F^{c}(0)$ and $P_F^{D,c}(0)$ the projectors of $\tilde %H(0)$
%and $H_D(0)$ for energies in $]-\infty, E_F]$ and such that 
%$\tilde E_k(2\pi)>E_F$ and $E_k^D(2\pi)>E_F$. Only states whose energy cross $%E_F$ are retained.

%\vskip 0.5cm
 
$$
\tilde Q_F={\rm Tr} \tilde P_{F,0}^{c}={\rm Tr} (\tilde P_{F,0}-\tilde P_{F,0}^{nc})
={\rm Ind}(\tilde P_{F,0};\tilde P_{F,0}^{nc})
\eqno(3.3)
$$
We introduce a smooth, monotone increasing function of time $\varphi(t)$, $0\leq t\leq T$, 
$\varphi(0)=0$ and $\varphi(T)=2\pi$, describing the 
adiabatic switching of a flux quantum through the axis of the cylinder.
Let $U_t$ be the unitary time evolution associated to the time dependent
 Hamiltonian $\tilde H(\varphi(t))$. From Lemma 3, as $t$ varies the spectral branches in $\Delta$
do not cross, and are monotone increasing.
  So an application of the adiabatic theorem [15] assures that
  $U_T \tilde P_{F,0}^{nc}U_T^{\dagger}$ tends to $\tilde P_{F, 2\pi}$.
Thus there exists some large enough $T_0$ such that for 
$T>T_0$, the pair of projections $(\tilde P_{F,0}^{nc}; 
U_T^\dagger \tilde P_{F,2\pi} U_T)$ satisfies
$$
||\tilde P_{F,0}^{nc} - U_T^\dagger \tilde P_{F,2\pi} U_T || < 1 
\eqno(3.4)
$$
Thus  it is Fredholm and ${\rm Ind}(\tilde P_{F,0}^{nc};U_T^\dagger \tilde P_{F,2\pi} U_T)=0$.
Since  $\tilde P_{F,0}-\tilde P_{F,0}^{nc}$ is finite rank we can apply (3.1) to get
$$
\eqalign{
\tilde Q_F & =  {\rm Ind}(\tilde P_{F,0};\tilde P_{F,0}^{nc})
\cr & = {\rm Ind}(\tilde P_{F,0};U_T^\dagger\tilde P_{F,2\pi}U_T)
+ {\rm Ind}(U_T^\dagger\tilde P_{F,2\pi}U_T; \tilde P_{F,0}^{nc})
\cr & = {\rm Ind}(\tilde P_{F,0};U_T^\dagger\tilde P_{F,2\pi}U_T)
\cr}
\eqno(3.5)
$$
Finaly let $U$ be the multiplication operator by $e^{i {2\pi\over L} y}$. Since
$U$ does not change the boundary conditions and $U^\dagger H(0)U=H(2\pi)$
we obtain the formula
$$
\tilde Q_F={\rm Ind}(\tilde P_{F,0}; U_T^\dagger U^\dagger \tilde P_{F,0} U U_T)
\eqno(3.6)
$$
The same construction for $H_D(0)$ leads to 
$$
Q_F^D={\rm Ind}(P_{F,0}^D; U_T^{D\dagger} U^\dagger P_{F,0}^D U U_T^D)
\eqno(3.7)
$$
where $P_{F,0}^D$ is the projector of $H_D(0)$ onto $]-\infty, E_F]$ and
 $U_t^D$ is the time evolution associated to the Hamiltonian $H_D(\varphi(t))$.
We remark that the identities of this paragraph
can be checked by explicit computation for the simple toy Hamiltonian (1.7).

\vskip 0.5cm 

\noindent{\bf Remark:} In [18] a different relative index for an infinite
two dimensional system is studied and related to the Hall conductivity viewed as a Chern number. It would be interesting to investigate the analogous
relationship in the present case with a boundary.

\vskip 0.5cm

\noindent{\bf Equality of $\tilde Q_F$ and $Q_F^D$.}

Since $V-V_D$ has a finite support $(z-H_D(0))^{-1}(V-V_D)$ is a compact 
operator for $z$ not in $\sigma(H_D(0))$. Therefore the resolvent identity 
and Cauchy's formula imply that $\tilde P_{F,0}-P_{F,0}^D$ is compact. Thus 
 the pair 
$(\tilde P_{F,0}; P_{F,0}^D)$ is Fredholm and we can apply (3.1) to get
$$
\eqalign{
{\rm Ind}(\tilde P_{F,0}; & U_T^\dagger U^\dagger\tilde P_{F,0}U U_T)
  \cr & = 
{\rm Ind}(\tilde P_{F,0}; P_{F,0}^D)+
{\rm Ind}( P_{F,0}^D;U_T^\dagger U^\dagger\tilde P_{F,0}U U_T)
\cr &
= 
{\rm Ind}(\tilde P_{F,0}; P_{F,0}^D)+
{\rm Ind}( P_{F,0}^D;U_T^\dagger U^\dagger P_{F,0}^DU U_T)
\cr &
+
{\rm Ind}(U_T^\dagger U^\dagger P_{F,0}^DU U_T;
U_T^\dagger U \tilde P_{F,0}U^\dagger U_T)
\cr}
\eqno(3.8)
$$
The first and third terms in the last equality of (3.8) cancel. Thus
$$
\eqalign{
\tilde Q_F & =
{\rm Ind}( P_{F,0}^D;U_T^\dagger U^\dagger P_{F,0}^DU U_T)
\cr &
= {\rm Ind}(P_{F,0}^D U U_T P_{F,0}^D| P_{F,0}^D{\cal H}\to
P_{F,0}^D{\cal H})
\cr}
\eqno(3.9)
$$
where in the last line we introduced the Fredholm index of 
$P_{F,0}^D U U_T P_{F,0}^D$ viewed as a map from
$P_{F,0}^D{\cal H}$ to itself (${\cal H}$ the Hilbert space of the cylinder).
From Dyson's equation
$$
P_{F,0}^D U U_T P_{F,0}^D - P_{F,0}^D U U_T^D P_{F,0}^D
= \int_0^T ds 
P_{F,0}^D U U_{T-s}^D(V-V_D)U_s P_{F,0}^D
\eqno(3.10)
$$
Therefore the Hilbert-Schmidt norm of the left hand side is smaller than
$$
\int_0^T ds ||P_{F,0}^D U U_{T-s}^D(V-V_D)||_{HS}
\eqno(3.11)
$$
which is shown to be finite in Appendix B. Thus the difference (3.10)
is compact and the two operators have the same Fredholm index
$$
\eqalign{
{\rm Ind}(P_{F,0}^D U U_T P_{F,0}^D & | P_{F,0}^D{\cal H}\to
P_{F,0}^D{\cal H})
\cr &
={\rm Ind}(P_{F,0}^D U U_T^D P_{F,0}^D| P_{F,0}^D{\cal H}\to
P_{F,0}^D{\cal H})
\cr}
\eqno(3.12)
$$
which is equivalent to $\tilde Q_F= Q_F^D$.

\vskip 0.5cm

\noindent {\bf End of Proof of (1.5) and (1.6).}

From the analysis of section 2 we know that for $D$ large enough
(say $D=O(L)$) the branches of $H_e(\Phi)$ and $H_D(\Phi)$ that belong to
$\Delta$ lie close to each other within a distance $O(e^{-c\sqrt B L})$. 
Since the spacing of the branches of $H_e(\Phi)$ is $O(L^{-1})$ it follows that
$Q_F^D= 1$ and therefore $\tilde Q_F=1$.
Thus $\tilde E_k(2\pi)=\tilde E_{k+1}(0)$ and since there exists $0\leq \bar\Phi\leq 2\pi$ such that
$$
\tilde E_k(2\pi)-\tilde E_k(0)=2\pi {d \tilde E_k\over d\Phi}(\bar\Phi)
\eqno(3.13)
$$
from (2.30) we get the lower bound
$$
| \tilde E_{k+1}(0)- \tilde E_k(0)|\geq {2\pi \tilde\alpha\over L}
\eqno(3.14)
$$
Because $\tilde E_l(0)=E_l(0)$, this bound shows that the levels of $H(0)$ 
(or $H(2\pi)$) are spaced by $O(L^{-1})$. Using the spectral flow of $\tilde H(\Phi)$, together with the facts that the levels of 
$\tilde H(\Phi)$ and $H(\Phi)$ are separated by $O(L^{-10})$, and that ${dE_k(\Phi)\over d\Phi}$is strictly positive, one deduces that
necessarily $E_k(2\pi)=E_{k+1}(0)$. Then proceeding as in (3.13) and (3.14)
we obtain the lower bound (1.6).
Finaly the upper bound is a consequence of
$$
\eqalign{
L|{dE_k\over d\Phi}(\bar\Phi)| & =<\Psi_k(\bar\Phi)|p_y-Bx+{\bar\Phi\over L}|
\Psi_k(\bar\Phi)>
\cr &
\leq
|| \Psi_k(\bar\Phi)||.||(p_y-Bx+{\bar\Phi\over L})\Psi_k(\bar\Phi)||
\cr &
\leq
(<\Psi_k(\bar\Phi)|2H(\bar\Phi)|
\Psi_k(\bar\Phi)> - <\Psi_k(\bar\Phi)|2V|
\Psi_k(\bar\Phi)>)^{1\over 2}
\cr &
\leq (2E_k(\bar\Phi)+2w)^{1\over 2}\leq (3B)^{1\over 2}
\cr}
\eqno(3.15)
$$

\vskip 0.5cm

\beginsection{APPENDIX A}

We start with a sketch of  preliminary estimates for the Green function of 
the  pure magnetic problem on the cylinder of circumference $L$, 
$$
H_0(\Phi)={1\over 2}p_x^2+{1\over 2}(p_y-Bx+{\Phi\over L})^2
\eqno(A.1)
$$
Using the spectral decomposition of the Green function 
$G_0(z)=(H_0(\Phi)-z)^{-1}$ on a basis of eigenfunctions 
$$
e^{i{2\pi m\over L}y }\varphi_{n,m}(x)
\eqno(A.2)
$$
and the Poisson summation formula we obtain
$$
<x,y|G_0(\Phi)|x^\prime, y^\prime>=
\sum_{m=-\infty}^{+\infty} e^{i{\Phi\over L}(y-y^\prime-mL)}
<x,y-mL|G_{0,\infty}(z)|x^\prime,y^\prime>
\eqno(A.3)
$$
where $G_{0,\infty}(z)$ is the Green function of the pure magnetic problem on the infinite
two dimensional plane. In the Landau gauge (${\bf r}=(x,y)$)
$$
<{\bf r}|G_{0,\infty}(z)|{\bf r^\prime}>
={B\over 2}\Gamma({1\over 2}-{z\over B})
U({1\over 2}-{z\over B}, 1, {B\over 2}|{\bf r}-{\bf r^\prime}|^2)
\exp(-{B\over 4}|{\bf r}-{\bf r^\prime}|^2+{iB\over 4}(x+x^\prime)(y-y^\prime))
\eqno(A.4)
$$
The presence of the Euler $\Gamma$ function indicates that the Landau levels remain unchanged
on the cylinder, and $U$ is the Kummer function [16]. By using some technical estimates 
as in [17] one may show that for ${B\over 2}<Re z<{3B\over 2}$ 
the absolute value of (A.3) is bounded above by
the simple expression
$$
{cB\over \delta_0(z)}e^{-{B\over 8}|x-x^\prime|^2}\sum_{m=-1,0,+1}
S(x-x^\prime, y-y^\prime-mL)e^{-{B\over 8}(y-y^\prime-mL)^2}
\eqno(A.5)
$$
where $c$ is a numerical constant independent of $B$ and $L$. The factor $S$ comes from the 
logarithmic divergence at coincident points
$$
\eqalign{
S(x-x^\prime, y-y^\prime)&=1 \qquad {\rm for} \qquad {B\over 2}|{\bf r}-{\bf r^\prime}|^2>1
\cr &
=\ln {B\over 2}|{\bf r}-{\bf r^\prime}|^2 \qquad {\rm otherwise}
\cr}
\eqno(A.6)
$$
A  bound similar to (A.5) holds for $|\partial_x<{\bf r}|G_{0,\infty}(z)|{\bf r^\prime}>|$, 
with $cB$ replaced by $cB^{3\over 2}$ and $S$  replaced by ${|x-x^\prime|\over 
|{\bf r}-{\bf r^\prime}|^2}$ when ${B\over 2}|{\bf r}-{\bf r^\prime}|^2<1$. The important 
feature for the subsequent estimates is that all the above singularities are integrable. In what follows $c$ denotes a generic numerical positive constant.

\vskip 0.5cm

\noindent{\bf Estimate of $||K_e||$.}

From the resolvent identity
$$
K_e(z)={1\over 2}[p_x^2, J_e]G_0(z)\chi_e+{1\over 2}[p_x^2, J_e]G_0(z)WG_e(z)\chi_e
\eqno(A.7)
$$
Evaluating the commutator, and using $||G_e(z)||\leq \delta_e(z)^{-1}$
we find
$$
||K_e(z)||\leq {1\over 2}||J_e^{\prime\prime}G_{0}(z)\chi_e||+
||J_e^{\prime}\partial_xG_{0}(z)\chi_e|| +
\delta_e(z)^{-1}(||J_e^{\prime\prime}G_{0}(z)W||+
||J_e^{\prime}\partial_xG_{0}(z)W||)
\eqno(A.8)
$$
Estimate (2.7) follows from the fact that all norms on the right hand side  
of (A.8) involve
matrix elements of $G_0(z)$ and $\partial_x G_0(z)$ separated by a distance at least equal
to ${D\over 4}$. We use the estimate ($A$ an operator with kernel $A({\bf r}, {\bf r^\prime})$)
$$
||A||\leq {\rm max} \biggl(\sup_{\bf r^\prime}\int d{\bf r}|A({\bf r}, {\bf r^\prime})|; 
\sup_{\bf r}\int d{\bf r^\prime}|A({\bf r}, {\bf r^\prime})|\biggr)
\eqno(A.9)
$$
For the first norm we have
$$
\eqalign{
\int_{-{3D\over 4}-1}^{-{3D\over 4}+1} dx \int_{L\over 2}^{L\over 2} dy  
& J_e^{\prime\prime}(x) 
|<{\bf r}|G_0(z)|{\bf r^\prime}>|\chi_e(x^\prime)
\leq 
{cBL\over \delta_0(z)}\int_{-{3D\over 4}-1}^{-{3D\over 4}+1} dx
e^{-{B\over 8}|x-x^\prime|^2}\chi_e(x^\prime)\cr &
\leq 
{c \sqrt B L\over \delta_0(z)}e^{-cBD^2}
\cr}
\eqno(A.10)
$$
In the first inequality we used (A.5) and in the last one we use the fact that 
$|x-x^\prime|\geq {D\over 4}$. On the other hand
$$
\eqalign{
J_e^{\prime\prime}(x) & \int_{-{D\over 2}}^{\infty} dx^\prime\int_{-{L\over 2}}^{{L\over 2}} dy^\prime
|<{\bf r}|G_0(z)|{\bf r^\prime}>|\chi_e(x^\prime)\cr &
\leq 
{cBL\over \delta_0(z)} J_e^{\prime\prime}(x)  \int_{-{D\over 2}}^{\infty} dx^\prime
e^{-{B\over 8}|x-x^\prime|^2}\chi_e(x^\prime)
\leq {c\sqrt BL\over \delta_0(z)}e^{-cBD^2}
\cr}
\eqno(A.11)
$$
Thus $||J_e^{\prime\prime} G_0(z) \chi_e||\leq {CL^2\over \delta_0(z)}e^{-cBD^2}$. For the term 
involving $\partial_xG_0(z)$ the estimates are similar. The terms involving $W$ lead to the same estimates
provided  
$$
\int_{-{3D\over 4}-1}^{-{3D\over 4}+1}dx e^{-{B\over 8}|x-x^\prime|^2} U(x^\prime)
\qquad {\rm and} \qquad
J_e^{\prime\prime}(x)\int_{0}^{\infty} dx^\prime e^{-{B\over 8}|x-x^\prime|^2} U(x^\prime)
\eqno(A.12)
$$
are bounded by $O(\exp(-cBD^2))$. This is the case for the class of functions $W(x)$ that 
 grow polynomialy as $x\to +\infty$.

\vskip 0.5cm

\noindent{\bf Estimate for $||K_b||$}

First we sketch the derivation of an estimate for the kernel of $G_b(z)$ and its derivative
for $z$ in the gap of $\sigma(H_b(\Phi))$.
$$
\eqalign{
&<{\bf r}|G_b(z)|{\bf r^\prime}>=<{\bf r}|G_0(z)|{\bf r^\prime}>
+ \sum_{m\geq 1}\int {\bf dr_1}...\int{\bf dr_m}
<{\bf r}|G_0(z)|{\bf r_1}>V({\bf r_1})\cr&\times<{\bf r_1}|G_0(z)|{\bf r_2}>V({\bf r_2})
...V({\bf r_m})<{\bf r_m}|G_0(z)|{\bf r^\prime}>\cr}
\eqno(A.13)
$$
Here the range of the integrals over $x_1,...,x_m$ is $]-\infty, +\infty[$, and that of 
$y_1,...,y_m$ is $[-{L\over 2},{L\over 2}]$. In order to extract the decay for $|x-x^\prime|$ large from 
(A.13) and (A.5) we use, from $B|x-x^\prime|^2>2\sqrt B |x-x^\prime|-1$, 
$$
\eqalign{
 & e^{-{B\over 8}(|x-x_1|^2+|x_1-x_2|^2+...+|x_m-x^\prime|^2)}
 \leq 
e^{-{B\over 16}(|x-x_1|^2+|x_1-x_2|^2+...+|x_m-x^\prime|^2)}\cr &
\times e^{-{\sqrt B\over 8}(|x-x_1|+|x_1-x_2|+...+|x_m-x^\prime|)}
e^{m\over 16}
\cr & \leq e^{m\over 16}e^{-{\sqrt B\over 8}|x-x^\prime|}
e^{-{B\over 16}(|x-x_1|^2+|x_1-x_2|^2+...+|x_m-x^\prime|^2)}
\cr}
\eqno(A.14)
$$
Thanks to (A.5), (A.13), (A.14) we obtain for ${B\over 2}|x-x^\prime|>1$
$$
\eqalign{
|<{\bf r}|G_b(z)|{\bf r^\prime}>| & \leq {cB\over \delta_0(z)}e^{-{B\over 8}|x-x^\prime|^2}
+\sum_{m\geq 1} ({cB\over \delta_0(z)})^{m+1}({w\over B})^m
e^{-{\sqrt B\over 8}|x-x^\prime|}
\cr &
\leq
{cB\over \delta_0(z)-cw} e^{-{\sqrt B\over 8}|x-x^\prime|}
\cr}
\eqno(A.15)
$$
This bound is valid as long as $w$ is small enough. Clearly from (A.13), 
following the same steps, we obtain a similar inequality, with $cB$ replaced
by $cB^{3\over 2}$, for
$|\partial_x<{\bf r}|G_b(z)|{\bf r^\prime}>|$ if ${B\over 2}|x-x^\prime|>1$.

To estimate $||K_b||$ we have to compute the norms on the right hand side of 
$$
||K_b||\leq {1\over 2}||J_b^{\prime\prime}G_b(z)\chi_b||+||J_b^\prime\partial_xG_b(z)\chi_b||
\eqno(A.16)
$$
This can be done easily using (A.9), (A.16) and the bound (A.15) together with that on 
the derivative. Then one finds
$$
||K_b||\leq {cB^{3\over 2}L\over \delta_0(z)-cw} e^{-c\sqrt B D}
\eqno(A.17)
$$

\vskip 0.5cm

\beginsection{APPENDIX B}

By Cauchy's formula, and the resolvent identity 
$$
P_{F,0}^D=\int_{\Gamma_F} dz {1\over z-H_D(0)}=
\int_{\Gamma_F} dz {1\over z-H_0(0)}+\int_{\Gamma_F} dz {1\over z-H_D(0)}(W+V_D){1\over z-H_0(0)}
\eqno(B.1)
$$
where the contour $\Gamma_F$ encloses the part of the spectrum of $H_D(0)$ 
lying below $E_F$. Setting
$g=U U_{T-s}^D(V-V_D)$ we have for the Hilbert-Schmidt norm
$$
\eqalign{
&||P_{F,0}^Dg||_{HS}\leq |\Gamma_F|{\rm sup}_{z\in \Gamma_F}||{1\over z-H_0(0)}g||_{HS}
\cr& +{|\Gamma_F|\over {\rm dist}(E_F, \sigma(H_D(0)))}
\biggl({\rm sup}_{z\in \Gamma_F}||W{1\over z-H_0(0)}g||_{HS}
+w{\rm sup}_{z\in \Gamma_F}||{1\over z-H_0(0)}g||_{HS}\biggr)
\cr}
\eqno(B.2)
$$
Here $|\Gamma_F|$ is the length of the contour which is finite because the spectrum
 is bounded below. Since $V-V_D$ has compact support, $g$ is 
 a square integrable
function on the cylinder. Therefore from the bound (A.5), (A.6) on the kernel of 
$(z-H_0)^{-1}$ it is easily seen that all the Hilbert-Schmidt norms in (B.2) are finite. These norms can be bounded above uniformly in $0\leq s\leq T$, and 
the supremum over $z$ stays finite as long as the contour does not touch a Landau level. Therefore (3.11) is finite.

\vskip 0.5cm 

\noindent {\bf ACKNOWLEDGEMENTS.} I wish to thank J\"urg Fr\"ohlich for drawing my attention on the spectral flow.

\beginsection{REFERENCES}

\noindent

\noindent [1] R. B. Laughlin, "Quantized Hall conductivity in two dimensions", Phys. Rev. B{\bf 23},
5632-5633 (1981) 

\noindent [2] B. I. Halperin, "Quantized Hall conductance, current carrying edge states, and the existence of extended states in a two dimensional disordered
potential", Phys. Rev B{\bf 25}, 2185-2190 (1982)

\noindent 

\noindent [3] N. Macris, Ph. A. Martin, J. V. Pul\'e, "On edge states in semi-infinite quantum Hall systems", J. Phys. A {\bf 32}, 1985-1996 (1999)

\noindent

\noindent [4] J. Fr\"ohlich, G. M. Graf, J. Walcher, "On the extended nature of edge states of quantum Hall hamiltonians", Ann. H. Poincar\'e {\bf 1}, 405 (2000)

\noindent

\noindent [5] S. De Bievre, J. V. Pul\'e, "Propagating edge states for a magnetic hamiltonian", Elect. J. Math. Phys. {\bf 5} (1999); http://mpej.unige.ch/mpej/MPEJ.html 

\noindent

\noindent [6] N. Macris, Ch. Ferrari, preprint EPFL.

\noindent

\noindent [7] S. Molchanov, "The local structure of the spectrum of 
the one dimensional Schroedinger operator", Comm. Math. Phys {\bf 78}, 429-446 (1981)

\noindent

\noindent [8] N. Minami, "Local fluctuation of the spectrum of a multidimensional Anderson tight binding model", Comm. Math. Phys {\bf 177}, 709-725 (1996) 

\noindent

\noindent [9] B. I. Shklovskii, B. Shapiro, B. R. Sears, P. Lambrianides,
H. B. Shore, "Statistics of spectra of disordered systems near the metal
insulator transition", Phys. Rev. B {\bf 47} 11487-11490 (1993)

\noindent

\noindent [10] J. Kellendonk, T. Richter, H. Schulz-Baldes, "Edge versus bulk currents in the integer quantum Hall effect", J. Phys. A: Math Gen {\bf 33}, 27-32 
(2000); and "Edge current channels and Chern numbers in the integer quantum 
Hall effect", mp-arc/00-266 

\noindent

\noindent [11] T. Kato, "Perturbation theory of linear operators", Springer Verlag, Berlin (1980)

\noindent

\noindent [12] P. Briet, J. M. Combes, P. Duclos, "Spectral stability under tunneling", Comm. Math. Phys {\bf 1206}, 133 (1989)

\noindent 

\noindent [13] F. Bentosela, V. Grechi, "Stark Wannier ladders", Comm. Math. Phys {\bf 142}, 169 (1991)

\noindent

\noindent [14] J. E. Avron, R. Seiler, B. Simon, "The index of a pair of projections", J. Funct. Anal,  220-237 (1994)

\noindent

\noindent [15] A. Messiah "Quantum mechanics", volume II, North Holland publishing company, Amsterdam (1961)

\noindent

\noindent [16] M. Abramovitz, I. A. Stegun, 
"Handbook of mathematical functions", Dover Publications, 
New York (1965); see page  

\noindent

\noindent [17] T. Dorlas, N. Macris, J. V. Pul\'e, "Characterisation of the 
spectrum of the Landau hamiltonian with delta impurities", Comm. Math. Phys
{\bf 204}, 367-396 (1999)

\noindent

\noindent [18] J. E. Avron, R. Seiler, B. Simon, "Charge deficiency, charge transport and comparison of dimensions", Comm. Math. Phys. {\bf 159}, 399-422 (1994)

\bye